# SUSTAINABLE EVOLUTION IN AN EVER-CHANGING ENVIRONMENT: GENERAL CHARACTERIZATION

*Maria K. Koleva*

Institute of Catalysis, Bulgarian Academy of Sciences
1113 Sofia, BULGARIA
mkoleva@bas.bg

## ABSTRACT

*A complex interplay between the academic issue about generalization of the thermodynamics and the practical matter about setting standards for a sustainable evolution of both tailored devices and natural systems is considered. It is established that the measure for a sustainable evolution in an ever-changing environment appears as a Boltzmann-Gibbs weight. At the same time, this measure performs as a local thermodynamical potential which, at the expense of being released from the condition of entropy maximization, serves as grounds for a fundamental development of the idea of banning perpetuum mobile. It is proven that the best efficiency of each engine that operates reversibly never exceeds the efficiency of corresponding Carnot heat engine where the engine is free from necessity of a physical coupling to two heat reservoirs.*

## INTRODUCTION

The goal of modern nano- and informational science is to meet the rapidly growing demands of the industry in an energy-effective and non-extensive way. To this end, the issue whether the behavior of the enormous variety of tailored devices exhibits common characteristics is of primary importance for developing a general strategy aimed to help resolving particular tasks. So far, the scientific community has not well established opinion on this topic because it still lacks a concept that, along with predicting universality, preserves the specific-goal-oriented characteristics of a device.

At first glance the look for any universality seems a hopeless affair because the operation of each device is result of a highly specific complex interplay among the processes that governs its internal self-organization at the corresponding constraints set by the environment. So far the study of this interplay has been focused on its particularity: usually systems are considered coupled to mass and/or heat reservoirs with constant in the time characteristics. The enormous activity on this matter starts with the study of heat engines in the mid 19-th century and is still rapidly developing. Summarizing, it is firmly established that the operational characteristics of any device are determined by a specific for the system relation between the macroscopic variables of its self-organization (e.g. concentrations) and the current environmental constraints. The variety of these relations (equation-of-state) is enormous: it ranges from the ideal gas law (Clapeyron equation) to systems of coupled ordinary/partial differential equations for far-from-equilibrium reaction-diffusion systems. Yet, the sustainability of a given operational protocol requires not only smart technology but proving robustness of its functioning to small variations of the control parameters as well. Obviously, the matter of sustainability involves also biological and social systems which are commonly known to operate at permanent variability of the environment.

At this point we suggest that the notion of equation-of-state serves as appropriate common basis for studying the macroscopic behavior of both equilibrium and out-of-equilibrium systems. The reason is that while the equilibrium systems are characterized both by their energy function and the corresponding equation-of-state, the steady states of out-of-equilibrium systems are defined only by their dynamics, i.e. by their equations-of-state. A key advantage of this suggestion is that any universality obtained in the corresponding setting would turn equally available for both infinitely large systems and their nano-scale counterparts. At the same time, we expect that the specific properties of a system, size-effects included, are concentrated in a specific steady "archetype" whose properties are described by the corresponding equation-of-state.

The enormous variety of systems and interactions system-environment implies that any universality is to be derived on the grounds of a few weak constraints only. We claim that a single constraint turns sufficient and this is the constraint of boundedness. It implies that environmental variations succeed in arbitrary order but so that:
(i) their amplitude remains bounded within specific to the system margins;
(ii) the increments are also bounded – this constraint naturally arises from the condition that to stay stable, a system exchanges only finite amount of energy/matter/information with the environment;
(iii) the rate of their succession is to be bounded so that a system to be able to reach steady state at current control parameters.
Note that the boundedness depends neither on the dimensionality of a system nor on its symmetry – it is applied to each control parameter and to each dimension separately.

A very powerful result of Lindeberg [1] proves that every bounded irregular sequence (*BIS*) has finite expectation value and finite variance. This result sets our first task: to demonstrate that the response to environmental variations modeled by a *BIS* is decomposable to a specific for the system steady component whose properties are described by the corresponding equation-of-state, and a stochastic one which has universal characteristics. Further, we shall demonstrate that this decomposition is insensitive both to the nature of the system and the environment, and to the statistics of the environmental variations. The universality of the fluctuations induced by the environmental variations sets our second task: to estimate the measure for exceeding by fluctuations the safety margins of a device. To this end, we prove in sec.2 that it always appears in a universal form. We demonstrate that the obtained measure generalizes the notion of Gibbs measure so that it acquires two-fold meaning: it appears both as local thermodynamical potential and as probability for robustness to environmental fluctuations. Besides, it turns free from the requirement to maximize the entropy which makes it available for large variety of complex systems including systems that exhibit different forms of macroscopic self-organization (e.g. pattern formation and morphology), i.e. systems whose common property is violation of the entropy maximization as condition for thermodynamical equilibrium.

Further, the release from the condition of entropy maximization and the corresponding non-probabilistic definition of entropy makes possible to define the notion of information on the grounds of the specific properties of the archetype and without any need of probability description.

In sec. 3, we develop the idea of banning perpetuum mobile proving that the efficiency of any device never exceeds that of corresponding Carnot heat engine where the engine is free from necessity of a physical coupling to two heat reservoirs. Moreover, we prove that though one can maximize the efficiency of any engine by manipulating the parameters of the cycle, still the best efficiency never reaches 100%.

**1. RESPONSE TO ENVIRONMENTAL FLUCTUATIONS MODELED BY A BIS**

The task of present section is to prove that the response of any system to a fluctuating environment modeled by a *BIS* is decomposable to a steady specific component (archetype) whose properties are described by the corresponding equation-of-state and a stochastic one which has universal characteristics.

We model a fluctuating environment by a vector of control parameters each component of which is a *BIS*. Then, the most general form of an equation-of-state in a fluctuating environment reads:

$$\vec{F}\left(\vec{\lambda}, \vec{n}, \frac{\partial \vec{n}}{\partial t}, \frac{\partial \vec{n}}{\partial x_i}, \frac{\partial^2 \vec{n}}{\partial x_i \partial x_j}\right) = 0 \qquad (1)$$

where $\vec{\lambda} = \vec{\lambda}_0 + \delta\vec{\lambda}_{st}$ is vector of the control parameters separated according to Lindeberg theorem [1] into expectation value $\vec{\lambda}_0$ and stochastic component $\delta\vec{\lambda}_{st}$; $\vec{n}$ is the vector of the macroscopic variables associated with the organization of the system (e.g. concentrations); the derivatives follow their traditional meaning (time and space correspondingly).

Eq.(1) determines the system response $\vec{n}(\vec{\lambda}_0 + \delta\vec{\lambda}_{st})$ to environmental variations which appears as a stochastic sequence whose terms are specified by the solution of eq.(1) for each realization of $\delta\vec{\lambda}_{st}$. In general, the equation-of-state eq.(1) may be arbitrary (algebraic or differential; deterministic or stochastic; linear or non-linear) - in each particular case its concrete form is specified by the interaction between a given system and its environment. Yet, it is subject to the following common constraint: it must be such that the stochastic sequence $\vec{n}(\vec{\lambda}_0 + \delta\vec{\lambda}_{st})$ is also a *BIS*. This condition is necessary for meeting the natural requirement that a system stays stable if and only if a finite amount of energy/matter/information is exchanged with the environment in every moment. Therefore, the response $\vec{n}(\vec{\lambda}_0 + \delta\vec{\lambda}_{st})$ is also a *BIS* and hence it is decomposable to expectation value (steady component) and a stochastic component:

$$\vec{n}(\vec{\lambda}_0 + \delta\vec{\lambda}_{st}) = \vec{n}_{\exp} + \delta\vec{n}_{st} \qquad (2)$$

where $\vec{n}_{\exp}$ is the expectation value and $\delta\vec{n}_{st}$ is a zero-mean *BIS*.

It should be stressed that the expectation value $\vec{n}_{\exp}$ from eq.(2) is not equal to $\vec{n}(\vec{\lambda}_0)$! This happens because the non-linearity of the equation-of-state eq.(1) renders the Taylor expansion of the terms in the stochastic sequence $\vec{n}(\vec{\lambda}_0 + \delta\vec{\lambda}_{st})$ around $\vec{\lambda}_0$ inappropriate. Yet, since $\vec{n}(\vec{\lambda}_0 + \delta\vec{\lambda}_{st})$ is *BIS* each term of which is determined by eq.(1), the expectation value $\vec{n}_{\exp}$ can be presented as a solution of the equation-of-state eq.(1) but at shifted control parameters $\vec{\lambda}_s$:

$$\vec{F}\left(\vec{\lambda}_s, \vec{n}_{\exp}, \frac{\partial \vec{n}_{\exp}}{\partial t}, \frac{\partial \vec{n}_{\exp}}{\partial x_i}, \frac{\partial^2 \vec{n}_{\exp}}{\partial x_i \partial x_j}\right) = 0 \qquad (3)$$

The shift between $\vec{\lambda}_s$ and $\lambda_0$ is highly specific to the system and the control parameters.

Thus, eq.(3) appears as the equation-of-state for the steady component ( archetype) of the response. It is worth noting that since it preserves the structure and the properties of eq.(1), the archetype preserves the specific goal-oriented properties of the system considered. Yet, since at the same time eq.(3) can be arbitrary, subject only to boundedness, an appropriate choice of a system opens the door to realization of a great variety of behavioral patterns ranging from a stationary state to a limit cycle, a self-assembled structure, morphological organization etc.

Outlining, one concludes that under the very weak constraint of boundedness, the response of a system to a fluctuating environment, modeled by a *BIS*, is additively decomposable into a specific steady component whose properties are defined by the corresponding equation-of-state, and a stochastic one which has universal characteristics. The decomposition is a generic property of eq.(1) and is insensitive to the nature of the system, to the particularities of its interaction with the environment, and to the statistics of the environmental variations.

## 2. BEYOND SAFETY THRESHOLDS. GENERALIZED GIBBS MEASURE

Now we come to the question about the universality of the stochastic component. Recently, we provided systematic study on that matter [2, ch. 1,3] and have proved that indeed *BIS`s* have certain characteristics that appear as time series invariants. These findings will help us to answer the question about the measure for escaping the safety thresholds. The issue about escaping safety thresholds is provoked by the non-linearity of the response which renders possible that certain members of the stochastic sequence $\delta\tilde{n}_{st}$ exceed the safety thresholds. Intuitively, it seems that the solution is very simple: for any particular sequence one calculates the probability for $\delta\tilde{n}_{st}$ to exceed the corresponding safety threshold. Hence, no universality is to be expected. However, this approach suffers a serious drawback: it does not take into account the role of the dynamics of the environmental variations. Its non-triviality is best revealed by obtained in [2, &1.7] result that the state space of a *BIS* is a dense transitive set of periodic orbits which implies that the motion in the attractor is indecomposable. Put it in other words, the latter implies that the average probability for escape is a non-local event and thus it depends not only on the current amplitude of the environmental variations but on the dynamics of their succession as well. Next we demonstrate that the complex interplay between the local amplitude of the environmental variations and the non-locality of their dynamics renders the average probability for exceeding safety thresholds to be described in a universal form. Further, it turns out that the non-locality of the dynamics is explicitly related to the heat dissipated by fluctuations and thus provides its appearing as an effective temperature in the expression for the average probability for escape. On the other hand, the boundedness imposed on the amplitude of fluctuations makes available to consider the orbital motion as confinement by a potential. We prove that the balance between the potential confinement and the random motion that comes from the dynamics of the fluctuations renders the average probability for escape the safety thresholds to appear in a universal form which generalizes the notion of the Boltzmann-Gibbs weight. Next follows the proof.

The structure of a *BIS* state space viewed as a dense transitive set of periodic orbits implies that the motion in it is completed as a random walk of blobs superimposed on an orbit where blobs are orbits on finer scale. The orbital motion on a coarse-grained scale is due to boundedness imposed on the amplitude of the variations. Accordingly, it acts as a confinement by a potential whose characteristics depend on the characteristics of the given orbit. On the other hand, the motion on the finer scales is random and is specified by the corresponding diffusion coefficient. Assuming that the diffusion coefficient exists, a statement that we shall prove later, it is obvious that the stationary motion in the state space is provided by the following balance between the random motion on finer scales and the potential confinement of the orbits on a coarse-grained scale:

$$h\overline{D}_U \frac{\partial P}{\partial n} = -\frac{\partial U(n)}{\partial n} P \qquad (4)$$

where $P$ is the probability for escaping from an orbit whose "potential" is $U(n)$; the properties of $U(n)$ are derived by means of considering the orbit as a Fourier knot [3], $n$ is a point is the state space (*it is worth noting that the points in the state space represent the values of the macroscopic variables not positions and velocities of the species in the system!*); $\overline{D}_U$ is the average diffusion coefficient and $h$ is a constant. Eq.(4) is always one-dimensional because the boundedness is insensitive to the symmetry of the system and its state space, and its dimensionality. Then its solution reads:

$$P(n) \propto \exp\left(-\frac{U(n)}{h\overline{D}_U}\right) \qquad (5)$$

Next item is to demonstrate that $k\overline{D}_U$ has the meaning of an effective temperature. For this purpose let us first derive an explicit formula for the average diffusion coefficient of a *BIS* bounded by a threshold corresponding to potential $U(n)$. The definition of a diffusion coefficient reads:

$$\overline{D} = \lim_{t \to \infty} \frac{X^2(t)}{t} \qquad (6)$$

where $X(t)$ is a deviation of a trajectory from the expectation value. Recently, we have established [2, &3.4.2] that the deviations from the expectation value of a *BIS* are developed as excursions of certain amplitude and duration; the relation between them is set by a power law where the exponent is specific to the system:

$$X \propto \Delta^{\beta(\Delta)} \qquad (7)$$

where $X$ is the amplitude of an excursion; $\Delta$ is the duration of its development; $\beta(\Delta)$ is an exponent set by the particularities of fine scale dynamics.

Further in [2, &3.4.3], we have established that the distribution of excursions reads:

$$G(X) = cX^{1/\beta(X)} \frac{\exp(-X^2/\sigma_U^2)}{\sigma_U} \tag{8}$$

where $\sigma_U$ is the variance of the sequence.

Then, eq.(6) becomes:

$$\overline{D}_U = \lim_{t \to \infty} \frac{\int_0^{X_U} X^2 G(X)}{\int_0^{\Delta_U} \Delta G(\Delta)} \tag{9}$$

By means of simple calculations $\overline{D}_U$ becomes:

$$\overline{D}_U \propto \frac{(\sigma_U)^2}{(X_U)^\alpha} \tag{10}$$

where $\alpha \ll 1$. Thus, eq.(10) proves that the diffusion coefficient of the random walk in the state space of a *BIS* is proportional to its variance. On the other hand, we have proved [2, &1.5] that the integrated power spectrum of a *BIS* is among its time series invariants and is also proportional to its variance. The non-triviality of this coincidence is revealed by the Nyquist theorem [4] which states that the integrated power spectrum is the measure for the heat dissipated in the interaction system-environment. Thus, one concludes that the diffusion coefficient $D_U$ is to be considered proportional to an effective temperature whose measure is the variance of the time series. Then, eq.(5) becomes:

$$P(n) \propto \exp\left(-\frac{U(n)}{kT_{eff}}\right) \tag{11}$$

It is worth noting that $P(n)$ from eq.(11) appears not only as measure for robustness to environmental variations but at the same time it serves as local thermodynamical potential for the given state of the system. More precisely, it appears as local Gibbs measure for a stationary bounded random motion. A crucial property of $P(n)$ is that it is not derived on the condition of entropy maximization but on the condition of balance between random and orbital motion given by eq.(4). To remind that in the traditional statistical mechanics the Boltzmann-Gibbs weight is that distribution whose entropy is maximal among all distributions that share the same fixed average energy [5]. It is worth noting, however, that this derivation is set by the explicit use of energy function as the basic characteristic of a system. However, as already have been stated in the Introduction, the energy function is not appropriate for characterizing the out-of-equilibrium systems. In order to overcome this difficulty, our derivation of the Boltzmann-Gibbs measure is set on the grounds of the equation-of-state and thus it turns equally available for both equilibrium and out-of-equilibrium systems. Its major advantage is that that now it is available not only for simple physical systems but it is available also for    systems that involve macroscopic self-organization including pattern formation, hierarchical organization etc, i.e. for systems that apparently violate the entropy maximization as a condition for equilibrium.

Accordingly, the notion of entropy acquires novel understanding. The above considerations open the door for defining the entropy as specific for a system non-probabilistic thermodynamic function that describes the amount of energy necessary for adapting the self-organization of that system to ever-changing control parameters. Indeed, the presence of a thermodynamical potential (Gibbs measure) renders the explicit mathematical expression of the entropy to be defined on the grounds of the Maxwell relations [6] so that its specification for each particular system is rendered by the corresponding equation-of-state. Thus, the present approach eliminates the major obstacle to the development of the notion of entropy beyond the simple physical systems, that of entropy maximization.

Further, it turns out that  the present approach provides unified thermodynamical description not only of  both equilibrium and out-of-equilibrium systems regardless of their size and level of macroscopic self-organization but it also brings about a completely novel prospective on the notion of information. Certainly, the release from the condition of entropy maximization makes available the association of different logical units with the steady states that are carried out at different control parameters, and the non-probabilistic definition of entropy provides the information encapsulated in each logical unit to be characterized by the specific for each of those steady states properties. *This setting is fundamentally different from the Shannon probabilistic approach to information which, on the contrary, leaves the notion of information insensitive to the specific properties of the steady states and their dynamics.*

### 3. GENERALIZATION OF THE CARNOT EFFICIENCY

The above obtained generalization of the thermodynamics poses the question about the corresponding generalization of its traditional implements. A central issue of this matter is the idea of banning perpetuum mobile. Along with an academic curiosity, the interest in this topic is provoked by its enormous importance for setting a standard for the best practical efficiency of any newly tailored device.

Next we prove that the idea about the Carnot engine admits very powerful generalization in the following sense: *the efficiency of any engine that operates reversibly never exceeds the efficiency of corresponding Carnot heat engine where the engine is free from necessity of a physical coupling to two heat reservoirs!*

If exists, an effective Carnot heat engine must be derived under weakest possible constraints. That is why, we start the derivation on the grounds of the first law of the thermodynamics since, being the energy conservation law, it is universally available:

$$dU = \dot{T}dS + \dot{S}dT + \dot{\vec{F}}_i \bullet d\vec{\Lambda}_i + \dot{\vec{\Lambda}} \bullet d\vec{F} \tag{12}$$

where $T$ is the temperature, $S$ is the entropy; $\vec{F}$ are the thermodynamical forces and $\vec{\Lambda}$ are their conjugates whose values are set by the control parameters.

The only constraint imposed over eq.(12) is that of boundedness of each derivative. This constraint commences from the natural requirement that the system stays stable if only the rates of exchange energy/matter/information with the environment are bounded.

Let us now consider that the system exerts a reversible cycle in the control parameter space. The crucial point is that the boundedness of all derivatives renders correspondence between the projections of the cycle in the $(F_i, \Lambda_i)$ section and $(T, S)$ sections so that to ban any degeneration of a cycle in $(F_i, \Lambda_i)$ section into a line in $(T, S)$ section. The reason is obvious: any line in the section $(T, S)$ implies infinity of the derivatives which, however, straightforwardly contradicts the idea of boundedness. This correspondence sets topological equivalence between the projection of every cycle in the $(T, S)$ section and that rectangular whose area and perimeter best fits both its area $W$ and perimeter $L$ - see Fig.1.

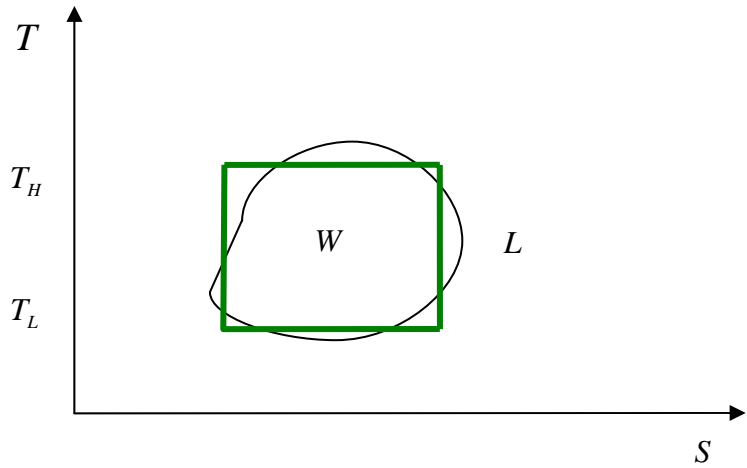

**Fig.1**

The importance of the fit by a rectangular is set by the circumstance that it presents the projection in the $(T, S)$ section of a Carnot cycle that operates reversibly between two effective "heat reservoirs" at "temperatures" $T_H$ and $T_L$. To remind that a Carnot cycle is a reversible thermodynamical cycle constituted by a sequence of isothermal and isentropic processes [7]; thus it is presented by a rectangular in the $(T, S)$ section.

Further, the perimeter of the original cycle $L$ also contributes to its energy balance because it is straightforwardly related to the energy dissipated by the fluctuations of the interaction system-environment. Indeed, this energy is proportional to the product of the perimeter $L$ and the integrated power spectrum of the fluctuations which as established in [2, &1.5] is proportional to the variance of the time series.

As a result, the approximation of the original cycle by an effective Carnot cycle makes available the evaluation of the efficiency of the tailored engine $\eta$ through the efficiency of the effective Carnot heat engine $\eta_0$:

$$\eta = \eta_0 = 1 - \frac{T_L}{T_H} \tag{13}$$

It should be stressed that the ban over degeneration of cycle in $(F_i, \Lambda_i)$ section into a line in $(T, S)$ section renders that the efficiency of the effective Carnot engine never to be zero and never to reach 100% !

It is worth noting that the above derivation is free from any dependence on either on the operational characteristics of the tailored engine or on the nature of its coupling to the environment. In turn, it fundamentally generalizes the idea of banning the perpetuum mobile because the release from the necessity of physical coupling to two heat reservoirs makes it available for every engine that operates reversibly. To this end, it should be stressed that though the efficiency of a tailored engine can be maximized by appropriate manipulating of the parameters of the cycle, the best efficiency never reaches 100%!

### CONCLUSIONS

The present paper demonstrates a highly non-trivial interplay between the academic issue about generalization of the thermodynamics and the practical matter about setting standards for a sustainable evolution of both tailored devices and natural systems in an ever-changing environment. The conjecture we put forward is to use the equation-of-state as the basic characteristic of a system. The advantage is that the equation-of-state is a common characteristics of both equilibrium and out-of-equilimbrium systems while the usually used energy function is appropriate only for equilibrium systems. Moreover, this setting is equally available for infinitely large systems and their nano-scale counterparts since any dependence of the dynamics on the size of a system appears explicitly only in the corresponding equation-of-state which sets the properties of the archetype.

It is proven that the above setting generalizes the notion of Boltzmann-Gibbs weight is the sense that under the condition of boundedness, it appears both as measure for robustness to environmental variations and as a local thermodynamical potential. Further, at the expense of being released from the condition of maximizing the entropy, the obtained measure serves as grounds for a fundamental development of the idea of banning perpetuum mobile. It is proven that the efficiency of any engine that operates reversibly never exceeds the efficiency of corresponding Carnot heat engine where the engine is free from necessity of a physical coupling to two heat reservoirs. It should be stressed that though the efficiency of a tailored engine can be maximized by appropriate manipulating of the parameters of the cycle, the best efficiency never reaches 100%!

Further, the new approach makes available to define the notion of information in a non-probabilistic way so that thus defined information is characterized by the specific properties of the archetype. In turn, thus obtained relation between the information and the specific organization of the system that creates it constitutes a fundamentally novel approach to the information theory, an approach that is very different from the probabilistic arguments of the traditional Shannon approach which renders the information and its processing insensitive to the specific properties of the system.

**REFERENCES**

1. W. Feller, An Introduction to Probability Theory and its Applications, J. Wiley&Sons; New York (1970)
2. M. K. Koleva, arXiv.org: physics/0512078
3. L. Kauffman, http://www.math.uic.edu/~kauffman/Fourier.html
4. C.W. Gardiner, Handbook of Stochastic Methods for Physics, Chemistry and Natural Sciences, 2-nd Edition, Springer; Berlin, Heidelberg (1985)
5. Maximum entropy probability distribution, wikipedia
6. Maxwell relations, wikipedia
7. Carnot heat engine, wikipedia